# DSA: Scalable Distributed Sequence Alignment System Using SIMD Instructions

Bo Xu, Changlong Li, Hang Zhuang, Jiali Wang, Qingfeng Wang, Jinhong Zhou, Xuehai Zhou
School of Computer Science and Technology
University of Science and Technology of China
Hefei, China
{xubo245, liclong, zhuangh, ustcwjl, qfwangyy, zjh}@mail.ustc.edu.cn, xhzhou@ustc.edu.cn

*Abstract*—Sequence alignment algorithms are a basic and critical component of many bioinformatics fields. With rapid development of sequencing technology, the fast growing reference database volumes and longer length of query sequence become new challenges for sequence alignment. However, the algorithm is prohibitively high in terms of time and space complexity. In this paper, we present DSA, a scalable distributed sequence alignment system that employs Spark to process sequences data in a horizontally scalable distributed environment, and leverages data parallel strategy based on Single Instruction Multiple Data (SIMD) instruction to parallelize the algorithm in each core of worker node. The experimental results demonstrate that 1) DSA has outstanding performance and achieves up to 201x speedup over SparkSW. 2) DSA has excellent scalability and achieves near linear speedup when increasing the number of nodes in cluster.

## I. INTRODUCTION

Sequence alignment algorithms are a basic and critical component of many bioinformatics fields for identifying regions of high similarity between pairwise sequences, such as read mapping, protein database searching and variation detection. With the development of next-generation sequencing (NGS) technology, the cost of sequencing drop faster than Moore's law, and the number of newly sequenced genomes has exhibited an exponential increase in recent years[1]. The fast growing reference database volumes and longer length of query sequence become new challenges for sequence alignment, so faster implementations of the sequence alignment algorithms are needed to keep pace.

The Smith-Waterman (SW) algorithm[2], which produces the optimal local alignment between query and reference sequences, is the most influential sequence alignment algorithm. However, it is also one of the slowest sequence alignment algorithms due to its computing and memory requirements grow quadratically in terms of the length of sequences. Heuristic approaches like BLAST[3] are considerably faster, but are not guaranteed to discover the optimal alignment.

Due to the critical role of SW, many efforts have been made to accelerate it, taking the advantages of special hardware such as single instruction multiple data (SIMD)[4-6], graphics processing unit (GPU)[7] and field programmable gate arrays (FPGA)[8]. Among these techniques, SIMD-based algorithms are most frequently used because they are compatible with most modern x86 CPUs[5]. Farrar[6] presented a new SW implementation where the SIMD registers are parallel to the query sequence, but are accessed in a striped pattern. Farrar's algorithm has been embedded in several popular genomic sequence mapping tools, such as BWA-MEM[9] and Bowtie2[10]. Zhao[5] extended Farrar's algorithm to return alignment information in addition to the optimal SW score. Daily[4] presented Parasail, a C-based library containing implementations of different pairwise sequence alignment algorithms by different instruction sets: SSE2, SSE4.1, AVX2 and KNC. These implementations take advantage of parallelization strategies. However, they show limited scalability.

To process bigger sequence data, it requires scalable storage and analysis frameworks. Adam[11] is genomic data processing system using Spark[12]. SparkSW[13] implements the SW algorithm on Spark for the first time and thus is load-balancing and scalable. However it only supports SW algorithm without the mapping location and traceback of the optimal alignment, which is quite important for read mapping and variation analysis. As a result, SparkSW is slow.

In order to accelerate sequence alignment in bigger sequence data and facilitate easy integration of scalable distributed sequence alignment algorithm into third-party software, this paper presents a scalable distributed sequence alignment system that uses SIMD instructions on Spark, which is called DSA. DSA not only has excellent scalability, but also achieves outstanding performance. The major contributions of this paper are as follows:

- We design and implement a scalable distributed sequence alignment system that employs Spark to process sequences data in a horizontally scalable distributed environment, and leverages data parallel strategy based on SIMD instruction to parallelize the algorithm in each core of worker node.
- We employ memory-based distributed file system as primary storage, which speeds up I/O performance by serving data from memory in local node rather than disks, and reduces network traffic between nodes by caching hot files in memory.
- We provide application programming interfaces for third-party software, including distributed local, global, and semi-global sequence alignment algorithms.

The rest of this paper is organized as follows. Section II describes the design of DSA. In section III, we describe our experiments and performance evaluation. Section IV presents conclusion and future work.

## II. DESIGN OVERVIEW

This section overviews the design of DSA, and describes its architecture, workflow and API respectively.

### A. System Architecture

DSA employs a standard master-slave architecture (see Figure 1). Master is primarily responsible for managing metadata and cluster. Each worker consists of two layers: The first layer is storage layer. In order to speed up read and write, we employ a memory-based distributed file system as primary storage component instead of the conventional disk-based distributed file system. In DSA, HDFS is only used as persistence in storage layer. The second layer is compute layer. It is based on Spark, a memory-based distributed computing framework. Due to SIMD instruction can be used to operate registers in parallel and perform the same operation on multiple data points simultaneously, DSA leverages data parallel strategy based on SIMD instruction to parallelize the algorithm in each core of worker node.

However, Spark cannot directly invoke SIMD-based sequence alignment algorithms which are written in C language. In order to integrate SIMD-based sequence alignment algorithms into Spark, we design a mediator. To be specific, firstly we implement multiple java classes to invoke SIMD-based sequence alignment algorithms separately by using Java Native Interface (JNI). Secondly, we implement scala class to invoke java class and transform java object to scala object. Hence, Spark applications can indirect invoke SIMD-based sequence alignment algorithms by the mediator in DSA.

Spark's scheduler applies delay scheduling[14] to schedule tasks, which has poor data locality and high network overhead when spark read file from HDFS and the execution time of tasks is long. So in DSA, we employ Alluxio[15], a memory-based distributed file system, as primary storage component, which speeds up I/O performance by serving data from memory in local node rather than disks, and reduces network traffic between nodes by caching hot files in memory.

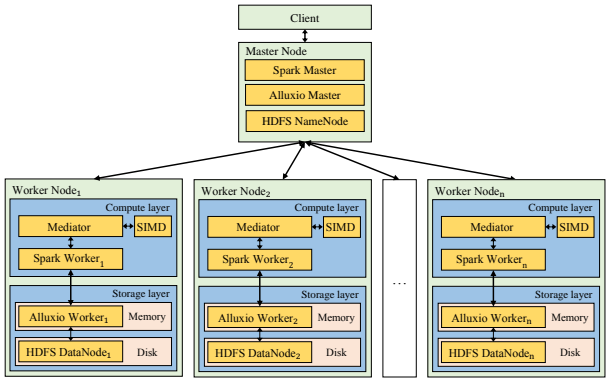

Figure 1. System architecture

### B. System Workflow

In process of sequence alignment, it is frequent to get the most K similar pairs of the alignments, such as protein database searching and seed extension of read mapping. Hence, DSA workflow consists of two main stages: map and top K phases (see Figure 2). Details of each phases are described as follows.

#### 1) Map

The map phase reads data directly from Alluxio in memory and creates RDD[12]. The input data of DSA include different datasets: query sequences, reference sequences and score matrix. The formats of query and reference sequences need to be suitable for distributed processing. DSA provides a converter for different formats.

After the data is ready, DSA uses SIMD technology to speed up sequence alignment in each map task. For local sequence alignment, DSA selects Striped Smith–Waterman(SSW)[5, 6] which was written for Intel processors supporting SSE2 instructions. For global and semi-global sequence alignment, DSA calls the Parasail library[4].

As shown in Figure 2, DSA will return a new RDD in map phase. There are many alignment results in each partition of the RDD and alignment results are alignment object including max score, the name of reference sequence, location, cigar and other properties.

#### 2) Top K

Once one map task has completed, DSA will execute top k algorithm. To be specific, DSA gets top k disordered alignment objects by using traditional quickselect algorithm separately in each partition of RDD. The select order is based on maxScore in alignment objects and implemented by an implicit method. Then DSA runs a reduce task. The reduce function aggregates global top k disordered alignment objects from all partitions. Finally, DSA sorts top k disordered alignment objects and returns results.

The time complexity of top k algorithm is $O(m + k * \log k + n * k)$, where $m$ is the number of alignment objects in a partition of RDD, $n$ is the number of partitions in RDD for simplicity of expression.

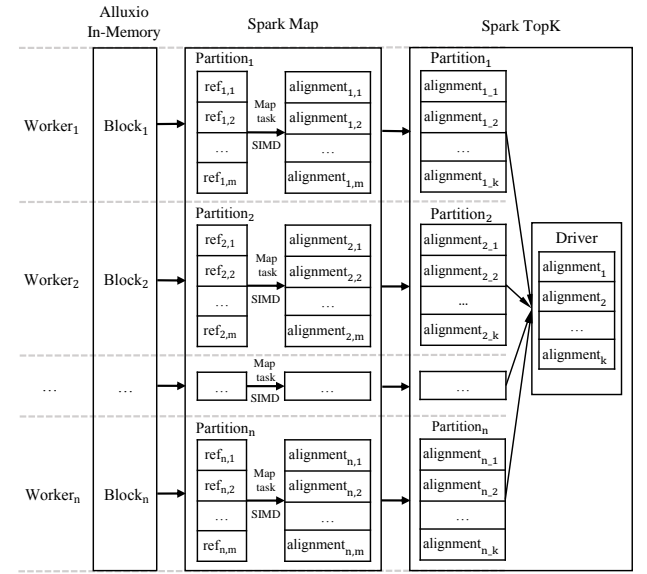

Figure 2. System workflow

## C. System API

In order to facilitate the users to use DSA directly and easy the integration of distributed sequence alignment algorithm for third-party software, we provide application programming interfaces (APIs) from two aspects as follows.

*1) Different data formats*

DSA provides a data preprocessing API implementation for different data formats, including FASTA, FASTQ and ADAM.

*2) Different algorithms*

DSA provides a distributed sequence alignment API implementation for different algorithms, including distributed Smith-Waterman (DSW), distributed Needleman-Wunsch (DNW) and distributed Semi-global (DSG) algorithms. DSA also allows user to define their own sequence alignment algorithm by extending a template class and override the implementation. It is important for third-party software to use their own algorithms in DSA.

## III. PERFORMANCE EVALUATION

In this section, we describe and evaluate DSA in two aspects: performance comparison with SparkSW, which is also distributed sequence alignment system on Spark, and the scalability of DSA is evaluated by a various number of computing nodes. The evaluation metric is speedup.

All our experiments were performed on a local cluster with 8 nodes. The operation system of each node is Ubuntu-14.04.1. Each node has a dual core Intel Xeon W3505 CPU with 22GB of RAM, and it is connected via Gigabit Ethernet. The Spark version is 1.5.2 and each node has 8GB executor memory for all Spark applications. The Alluxio version is 1.3.0 and each node has 12GB memory storage for Alluxio worker.

In order to achieve a better performance comparison with SparkSW, we use the same data as SparkSW[13]. The details of the experimental datasets are listed in Table I.

TABLE I. THE LIST OF REFERENCE AND QUERY SEQUENCES

| R | SR | NR | Q | NQ | LQ |
|---|----|----|---|----|----|
| R1 | 32 | 78295 | Q1 | P18691 | 8 |
| R2 | 64 | 156590 | Q2 | P83140 | 16 |
| R3 | 128 | 313180 | Q3 | P20738 | 32 |
| R4 | 256 | 626360 | Q4 | O55746 | 64 |
| R5 | 512 | 1252720 | Q5 | Q6GZW8 | 128 |
| R6 | 1024 | 2505440 | Q6 | Q6GZX4 | 256 |
| R7 | 2048 | 5019006 | Q7 | Q19LI2 | 512 |
| R8 | 4096 | 10038012 | Q8 | Q7TQI7 | 1024 |
| R9 | 8192 | 20076024 | Q9 | Q8IYD8 | 2048 |
| R10 | 16384 | 40152048 | Q10 | R0INU3 | 4096 |

R: order of reference; SR: size of reference (MB); NR: number of reference; Q: order of query; NQ: name of query; LQ: length of query (chars);

### A. Performance Comparison

The fast growing reference database volumes and longer length of query sequence are new challenges for sequence alignment. In this part, we design two different experiments to validate the ability of DSA as follows.

*1) Different size of reference databases*

The first experiment uses a query sequence with fixed length and different size of reference databases. The fixed length of query sequence is 512 chars (Q7), and the reference databases are R1 to R10 (see Table I). DSA and SparkSW both run distributed SW algorithm on a local cluster with 8 nodes and each node has 8GB executor memory for all Spark applications.

Figure 3 shows DSA's speedup over SparkSW in different size of reference databases. The experimental result shows that DSA has a significant performance improvement over SparkSW in different reference datasets. The maximum speedup is up to 122-fold.

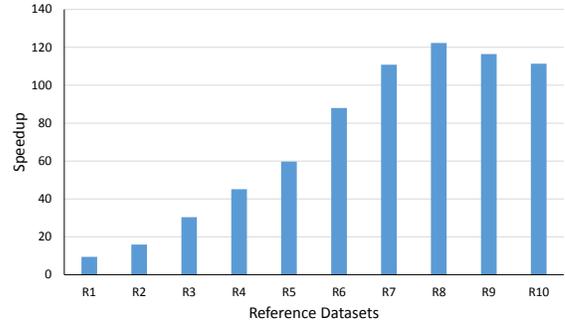

Figure 3. The comparison with SparkSW for fixed length of query and different size of reference database

*2) Different length of query sequences*

The second experiment uses a reference database with fixed size and different length of query sequences. The fixed size of reference database is 4G (D8) and the query sequences are Q1 to Q10 (see Table I). Other experimental environments are the same as previous experiment.

Figure 4 shows DSA's speedup over SparkSW in different length of query sequences. The experimental result shows that DSA has a significant performance improvement over SparkSW in different query datasets. The maximum speedup is up to 201-fold.

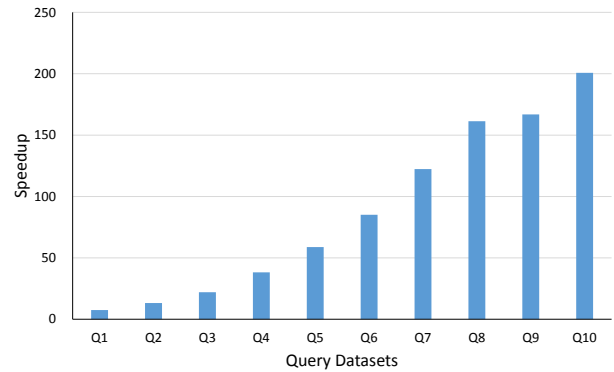

Figure 4. The comparison with SparkSW for fixed size of reference database and different length of query

We analyzed qualitatively why DSA is much faster than SparkSW. The reasons are shown as follows:

- The major reason is that DSA uses SIMD instructions, where the SIMD registers are parallel to the query sequence, but are accessed in a striped pattern.
- DSA leverages Alluxio instead of HDFS as primary storage, which speeds up I/O performance and reduces network traffic between nodes.
- DSA employs a more effective top k algorithm, which not only reduces the time complexity from $O(m*m + n*n)$ in SparkSW to $O(m + k*\log k + n*k)$, and k is usually relatively small, but also avoids shuffle like SparkSW.

Besides achieving higher performance, the accuracy of DSA is also improved on top of SparkSW because of the filtering strategy can filter some true results in open source SparkSW. Due to limited pages of poster, there is no detailed analysis in this paper.

*B. Evaluation of Scalability*

In order to evaluate the scalability of DSA, we run DSW, DNW and DSG algorithms of DSA with a various number of computing nodes. In this experiment, we select Q7 as query sequence and R8 as reference database (see Table I).

Figure 5 shows that different algorithms' speedup over themselves running on single node. The experimental result shows that the three algorithms of DSA achieve near linear speedup when increasing the number of nodes from 1 to 8 in cluster.

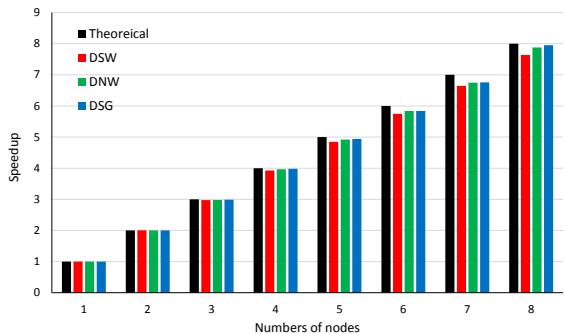

Figure 5. The speedup improvement by increasing the number of nodes

## IV. CONCLUSION AND FUTURE WORK

In this paper, we present and design DSA, a scalable distributed sequence alignment system that employs Spark to process sequences data in a horizontally scalable distributed environment, leverages data parallel strategy based on SIMD instruction to parallelize the algorithm in each core of worker node, and employs memory-based distributed file system Alluxio as primary storage to speeds up I/O performance and reduces network traffic. DSA not only provides a data preprocessing API implementation for different data formats, including FASTA, FASTQ and ADAM, but also provides a distributed sequence alignment API implementation for different algorithms, including distributed Smith-Waterman, Needleman-Wunsch and Semi-global alignment algorithms.

The experimental results demonstrate that DSA has outstanding performance and excellent scalability. DSA achieves up to 201x speedup over SparkSW and near linear speedup when increasing the number of nodes in cluster.

In the future, we plan to explore different Instruction sets to improve performance, and provide more API for different data formats and functions.

ACKNOWLEDGMENT

We would like to thank Bin Fan, who is PMC member of Alluxio. This work was supported by the National Science Foundation of China under grants No.61379040, No. 61272131 and No. 61202053.

AVAILABILITY

An open source DSA (GNU GPL v.2) are freely available at: https://github.com/xubo245/DSA.

REFERENCES

[1] P. Muir, S. Li, S. Lou, D. Wang, D. J. Spakowicz, L. Salichos, *et al.*, "The real cost of sequencing: scaling computation to keep pace with data generation," *Genome biology,* vol. 17, p. 1, 2016.
[2] T. F. Smith and M. S. Waterman, "Identification of common molecular subsequences," *Journal of molecular biology,* vol. 147, pp. 195-197, 1981.
[3] S. F. Altschul, T. L. Madden, A. A. Schäffer, J. Zhang, Z. Zhang, W. Miller, *et al.*, "Gapped BLAST and PSI-BLAST: a new generation of protein database search programs," *Nucleic acids research,* vol. 25, pp. 3389-3402, 1997.
[4] J. Daily, "Parasail: SIMD C library for global, semi-global, and local pairwise sequence alignments," *BMC bioinformatics,* vol. 17, p. 1, 2016.
[5] M. Zhao, W.-P. Lee, E. P. Garrison, and G. T. Marth, "SSW Library: an SIMD Smith-Waterman C/C++ library for use in genomic applications," *PloS one,* vol. 8, p. e82138, 2013.
[6] M. Farrar, "Striped Smith–Waterman speeds database searches six times over other SIMD implementations," *Bioinformatics,* vol. 23, pp. 156-161, 2007.
[7] Y. Liu, A. Wirawan, and B. Schmidt, "CUDASW++ 3.0: accelerating Smith-Waterman protein database search by coupling CPU and GPU SIMD instructions," *BMC Bioinformatics,* vol. 14, p.: 117., 2013.
[8] I. T. Li, W. Shum, and K. Truong, "160-fold acceleration of the Smith-Waterman algorithm using a field programmable gate array (FPGA)," *BMC bioinformatics,* vol. 8, p. 1, 2007.
[9] H. Li, "Aligning sequence reads, clone sequences and assembly contigs with BWA-MEM," *arXiv preprint arXiv:1303.3997,* 2013.
[10] B. Langmead and S. L. Salzberg, "Fast gapped-read alignment with Bowtie 2," *Nature methods,* vol. 9, pp. 357-359, 2012.
[11] M. Massie, F. Nothaft, C. Hartl, C. Kozanitis, A. Schumacher, A. D. Joseph, *et al.*, "Adam: Genomics formats and processing patterns for cloud scale computing," *EECS Department, University of California, Berkeley, Tech. Rep. UCB/EECS-2013-207,* 2013.
[12] M. Zaharia, M. Chowdhury, M. J. Franklin, S. Shenker, and I. Stoica, "Spark: cluster computing with working sets," *HotCloud,* vol. 10, pp. 10-10, 2010.
[13] G. Zhao, C. Ling, and D. Sun, "SparkSW: scalable distributed computing system for large-scale biological sequence alignment," in *Cluster, Cloud and Grid Computing (CCGrid), 2015 15th IEEE/ACM International Symposium on*, 2015, pp. 845-852.
[14] M. Zaharia, D. Borthakur, J. Sen Sarma, K. Elmeleegy, S. Shenker, and I. Stoica, "Delay scheduling: a simple technique for achieving locality and fairness in cluster scheduling," in *Proceedings of the 5th European conference on Computer systems*, 2010, pp. 265-278.
[15] H. Li, A. Ghodsi, M. Zaharia, S. Shenker, and I. Stoica, "Tachyon: Reliable, memory speed storage for cluster computing frameworks," in *Proceedings of the ACM Symposium on Cloud Computing*, 2014, pp. 1-15.